\documentclass{webofc}
\usepackage[varg]{txfonts}   
\usepackage[nameinlink,capitalize]{cleveref}

\usepackage{color}

\begin{document}
\title{Spectral reconstruction details of a gradient-flowed color-electric correlator}
%
%

\author{\firstname{Luis} \lastname{Altenkort}\inst{1}\fnsep\thanks{Speaker, \email{altenkort@physik.uni-bielefeld.de}} \and
\firstname{Alexander M.} \lastname{Eller}\inst{2} \and
        \firstname{Olaf} \lastname{Kaczmarek}\inst{1}  \and
        \firstname{Lukas} \lastname{Mazur}\inst{3} \and
        \firstname{Guy D.} \lastname{Moore}\inst{2} \and
        \firstname{Hai-Tao} \lastname{Shu}\inst{4}
}

\institute{Fakult\"at f\"ur Physik, Universit\"at Bielefeld, D-33615 Bielefeld,
    Germany
\and
Institut f\"ur Kernphysik, Technische Universit\"at Darmstadt, D-64289 Darmstadt, Germany
\and
            Paderborn Center for Parallel Computing, Paderborn University, D-33098 Paderborn, Germany
\and
          Institut f\"ur Theoretische Physik, Universit\"at  Regensburg, D-93040 Regensburg, Germany
          }

\abstract{%
In a recently published work we provide a proof-of-concept of a novel method to extract the heavy quark momentum diffusion coefficient from color-electric correlators on the lattice using gradient flow. The transport coefficient can be found in the infrared limit of the corresponding spectral function which is reconstructed through perturbative model fits of the correlator data. In this proceedings report we want to give more detailed insights into the systematic uncertainties of this procedure and compare our results with other studies.
}
\maketitle

\section{Introduction}
\label{sec-1}
Experimental data from heavy ion collisions provide evidence for a significant collective motion of heavy quarks (charm, bottom) with the bulk of the medium \cite{ALICE2018v2,ALICE2018RAA}.  From a hydrodynamical standpoint this is unexpected, as the kinetic equilibration time $\tau_\mathrm{kin}$ for heavy participants of the evolution should be about a factor of $M/T$ larger than that of the light ones \cite{Moore_2005}.
In a hot medium with temperature $T$ the behavior of heavy quarks (HQ) with mass $M$ can be described by a nonrelativistic Langevin approach, as long as $M\gg \pi T$  \cite{Moore_2005, BERAUDO200959}. From this description multiple a priori unknown coefficients emerge with direct phenomenological interpretations. One of these coefficients is the HQ momentum diffusion coefficient $\kappa$, which quantifies the averaged squared momentum transfer that is exerted on a heavy quark by the hot medium per unit time. In the nonrelativistic limit it is directly related to other coefficients such as the spatial diffusion coefficient $D$ and the drag coefficient $\eta_D$, as well as to the kinetic equilibration time $\tau_\mathrm{kin}$. In this way a determination of $\kappa$ also determines $\tau_\mathrm{kin}$. However, even for high temperatures perturbative series for $D$ or $\kappa$ are ill-behaved \cite{Caron_Huot_2008} and a first principles determination from lattice QCD is necessary to provide insight into the observed behavior of heavy quarks. The coefficient $\kappa$ is also a crucial input for simulation models \cite{Brambilla:2020qwo}.

Diffusion physics is encoded in the infrared limit of real-time spectral functions of conserved currents \cite{kubo,hydrodynamicfluctuations,Meyer2011}. In the limit $M \gg \pi T$ one can construct a color-electric correlator whose spectral function encodes $\kappa$ through \cite{CaronHuot:2009uh}
\begin{align}
    \kappa = \lim_{\omega\rightarrow 0} 2T \frac{{{\rho}{(\omega)}}}{\omega},  \quad
    G(\tau) = \int^\infty_0 \frac{\mathrm{d} \omega}{\pi} \,  \frac{\cosh(\omega(\tau-\beta/2))}{\sinh(\omega\beta/2)} \, {{\rho}{(\omega)} }.
    \quad \beta=1/T.
    \label{eq:kappa_and_spfintegral}
\end{align}
On a Euclidean space-time lattice the spectral function $\rho(\omega)$ is not directly accessible; instead on has to invert the above integral relation using the Euclidean correlation function $G(\tau)$.
This is an ill-posed problem as it is indeterminate by definition; nevertheless it is possible to get meaningful results from this approach through various ameliorating techniques. 
$G(\tau$) is a purely gluonic color-electric correlation function of Euclidean time $\tau$ that can be constructed by utilizing Heavy Quark Effective Theory \cite{CaronHuot:2009uh}. It reads
\begin{align}
    G(\tau) = -\frac{1}{3} \sum_{i=1}^3 \frac{\left\langle \operatorname{Re} \left[ \mathrm{tr} \left[U{(\beta, \tau)} \enspace g E_i{ (\mathbf{0},\tau)} \enspace U{(\tau,0)} \enspace gE_i{ (\mathbf{0},0)}  \right] \right]\right\rangle}{\left\langle  \operatorname{Re} \left[ \mathrm{tr} [\, U{ (\beta,0)} \,] \right] \right\rangle},
    \label{eq:corr}
\end{align}
where $U(b,a)$ is a temporal Wilson line connecting $a$ and $b$, $E_i$ is the color-electric field operator and $g$ is the coupling. The main advantage of using this gluonic color-electric correlator instead of genuine hadronic correlators is that its spectral function is expected to have a smooth infrared limit which simplifies the spectral reconstruction significantly. 



\section{Spectral reconstruction}

\begin{figure}[t]
\centering
\includegraphics[width=0.49\textwidth]{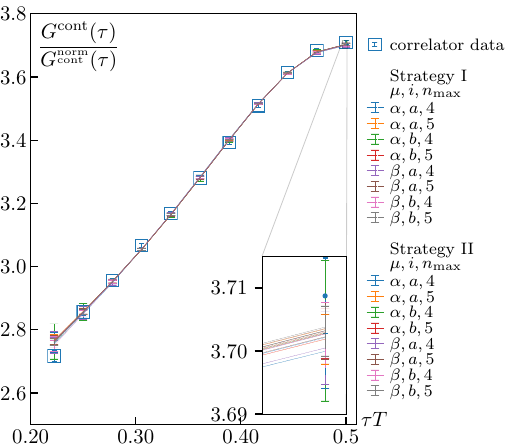}
\hfill \includegraphics[width=0.49\textwidth]{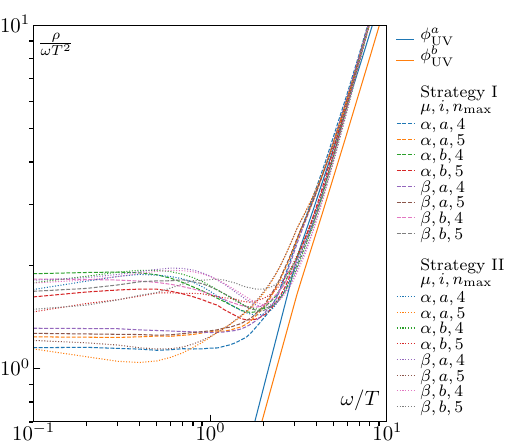}
\caption{Left: Model correlators obtained from  integration of the model spectral functions. The labels distinguish between different models (see \cref{eq:modelspf}). Right: Model spectral functions from fits of the correlator data for all considered models. Statistical errors are hidden for better visibility; for $\omega = 0$ they can be found in \cref{fig:kappa-D}, all others can be found in the data publication \cite{datapublication}.}
\label{fig:corr-spf-models}      
\end{figure}

One way to make the inversion of \cref{eq:kappa_and_spfintegral} less indeterminate is to constrain the form of the spectral function based on additional theoretical knowledge. The idea is to build a model spectral function that obeys the correct functional form in the low- and high-frequency limits with a general interpolation in between. 
The true spectral function of \cref{eq:kappa_and_spfintegral} is then substituted with this general model that has a finite number of parameters.  
Given some set of these parameters, one can perform the integral over the model spectral function and obtain the corresponding model correlator, $G^\textrm{model}$, which can be compared to the original correlator data, $G^{\rm{cont}}$, and its statistical error, $\delta G^{\rm{cont}}$. In this way one can define a weighted sum of squares through
\begin{align}
    \chi^2\equiv\sum_{\tau}\bigg{[}\frac{G^{\rm{cont}}(\tau)-G^\textrm{model}(\tau)}{\delta G^{\rm{cont}}(\tau)}\bigg{]}^2.
\label{eq:chisq}
\end{align}
The task is then to fit the parameters of the model spectral function such that \cref{eq:chisq} is minimized. If the model for the spectral function was appropriate, then the model correlator should be in agreement with the correlator data. Note that this approach is only reasonable if the spectral function is expected to be devoid of intricate structures or peaks, which is a reasonable assumption in this case \cite{CaronHuot:2009uh}. 
 
The spectral function of the color-electric correlator (\cref{eq:corr}) can be modeled through 
\begin{equation} 
     \rho_{\rm{model}}^{(\mu, i)}(\omega) \equiv \Big{[} 1 + \sum_{n=1}^{n_{\rm{max}}} c_n e^{(\mu)}_n(y) \Big{]} \sqrt{ \big{[} \phi_{\rm{IR}}(\omega)\big{]}^2 + \big{[} \phi^{(i)}_{\rm{UV}}(\omega)\big{]}^2 },
     \label{eq:modelspf}
\end{equation}
where the term in brackets is responsible for the interpolation (see \cite{Altenkort2021} for notation and details) between the infrared part $\phi_{\rm{IR}}$ and ultraviolet part $\phi_{\rm{UV}}$. For the interpolation and the UV part, different functional forms are reasonable; the superscripts $\mu \in \lbrace \alpha, \beta \rbrace$, $i \in \lbrace a , b \rbrace$ and $n_\mathrm{max} \in 4,5$ depict these (see \cite{Altenkort2021} for detailed explanations). The infrared part is modeled by the simplest functional form that is consistent with \cref{eq:kappa_and_spfintegral}, which reads $\phi_{\rm{IR}}(\omega) \equiv \frac{\kappa \omega}{2T}$ \cite{Francis:2015daa}. In the end we normalize with the temperature and so $\kappa/T^3$ is simply one of the free parameters. For the fit we follow two different strategies, which are explained in detail in \cite{Altenkort2021}. 
In summary we obtain 16 equally reasonable models that are distinguished by the parameters $(\mu,i)$, $n_\mathrm{max}$ and two fit strategies. 

\section{Results and comparison}\label{sec:results}

\begin{figure}[t]
\centering
\includegraphics[width=0.49\textwidth]{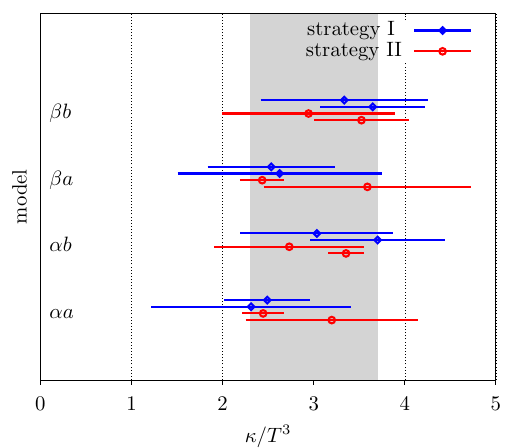}
\hfill\includegraphics[width=0.49\textwidth]{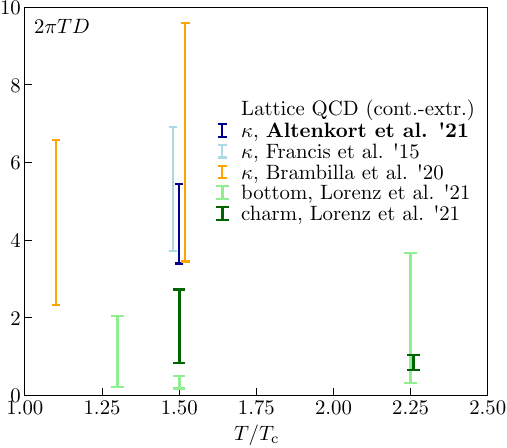}
\caption{Left: Heavy quark momentum diffusion coefficient $\kappa/T^3$ from all considered spectral function model fits. The grey band indicates the final range we cite (see \cref{sec:results}). Right: Comparison of the heavy quark spatial diffusion coefficient $2\pi TD$ from different studies \cite{Altenkort2021, Brambilla:2020, Francis:2015daa, Lorenz:2021}. Results from the heavy quark limit using the color-electric correlator (``$\kappa$'' in legend) are converted using $D=2T^2/\kappa$. Also shown are recent results obtained from vector meson correlators for charm and bottom quarks. }
\label{fig:kappa-D}       
\end{figure}

All our theoretically motivated models yield model correlators that agree with the original correlator data as can be seen from the left panel of \cref{fig:corr-spf-models}. 
The $\chi^2/\mathrm{d.o.f.}$ of the fits lie in the range $3.4 \dots 5.4$. On the right panel of \cref{fig:corr-spf-models} the resulting spectral functions are shown. 
Here, the estimate for the momentum diffusion coefficient $\kappa$ for each spectral function can be read off at the intercept with the y-axis. All of the intercepts are shown in the left panel of \cref{fig:kappa-D} with a grey band that indicates the final range we cite for the HQ momentum diffusion coefficient, which is $\kappa/T^3=2.31\,\dots\,3.70$. In the nonrelativistic limit ($M\gg \pi T$) we can use the Einstein relation $D=2T^2/\kappa$ to convert this to $2\pi TD=3.40\,\dots\,5.44$. We can also convert to the kinetic equilibration time $ \tau_\mathrm{kin}= {\eta_D^{-1}} = (1.63\,\dots\,2.61) \left( {\frac{T_c}{T} }\right)^2 \left({\frac{M}{1.5 \mathrm{GeV}}}\right)\mathrm{fm/c}$. 
In comparison with light degrees of freedom of the medium (see, for example, \cite{Schlichting:2019abc}), the results suggest a rather fast thermalization, which seems to fit the experimental observations of a large collective motion of heavy quarks with the medium.

In the right panel of \cref{fig:kappa-D} we compare to other studies \cite{Brambilla:2020, Francis:2015daa, Lorenz:2021}. 
It agrees with both values obtained from previous studies that use the multi-level method (instead of gradient flow) and similar spectral reconstruction methods of the same continuum-extrapolated color-electric correlator obtained in the heavy quark limit. However, for all results from the heavy quark limit there is still some friction with recent results from hadronic correlators of charm and bottom quarks with physical masses \cite{Lorenz:2021}. Recently it has been worked out that the leading finite mass correction to the momentum diffusion coefficient is encoded in a color-magnetic correlator \cite{Bouttefeux:2020ycy}, which can be computed using a similar strategy. We expect this correction to loosen the tension between the two methods; however, there are still open questions in the renormalization of the color-magnetic correlator that need to be solved first \cite{Laine:2021uzs}.

\section{Acknowledgements}
The authors acknowledge support by the Deutsche Forschungsgemeinschaft (DFG, German Research Foundation) through the CRC-TR 211 'Strong-interaction matter under extreme conditions'– project number 315477589 – TRR 211. The computations in this work were performed on the GPU cluster at Bielefeld University.

\end{document}